\documentclass[a4paper,traditabstract]{aa}
\usepackage{txfonts,natbib,graphicx,float,subfigure}
\bibpunct{(}{)}{;}{a}{}{,}

\begin{document}

\title{A very luminous, highly extinguished, very fast nova - V1721~Aquilae}
\author{R. Hounsell 
\inst{1}\fnmsep\thanks{\email{rah@astro.livjm.ac.uk}}
\and M. J. Darnley
\inst{1}
\and M. F. Bode
\inst{1}
\and D. J. Harman
\inst{1}
\and L. A. Helton 
\inst{2,}
\inst{3}
\and G. J. Schwarz
\inst{4}
\institute{Astrophysics Research Institute, Liverpool John Moores University, Twelve Quays House, Egerton Wharf, Birkenhead, CH41 1LD, UK.
\and SOFIA Science Center, USRA, NASA Ames Research Center, M.S. N211-3, Moffett Field, CA 94035, USA.
\and Department of Astronomy, The University of Minnesota, 116 Church St. S.E. Minneapolis, MN 55455, USA.
\and American Astronomical Society, 2000 Florida Ave., NW Suite 400, Washington, DC 20009-1231, USA.
\\
}}
\date{Received date / Accepted date}

\abstract{Studies indicate that fast novae are primarily located within the plane of the Milky Way and slow novae are found within its bulge. Because of high interstellar extinction along the line of sight many novae lying close to the plane are missed and only the brightest seen. One nova lying very close to the Galactic plane is V1721~Aquilae, which was discovered in outburst on 2008 September 22.5 UT. Examination of spectra obtained 2.69 days after outburst revealed very high expansion velocities (FWHM of the H$\alpha$ emission $\approx$ 6450 km s$^{-1}$). In this paper we have used available pre- and post-outburst photometry and post-outburst spectroscopy to conclude that the object is a very fast, luminous, and highly extinguished ($A_{V} = 11.6 \pm 0.2$) nova system with an average ejection velocity of $\approx 3400$ km $\mathrm{s^{-1}}$. Pre-outburst \textit{near}-IR colours from the 2MASS point source catalogue indicate that at quiescence the object is similar to many quiescent classical novae and appears to have a main sequence/sub-giant secondary rather than a giant counterpart. Based on the speed of decline of the nova and its emission line profiles we hypothesise that the axis ratio of the nova ejecta is $\sim 1.4$ and that its inclination is such that the central binary accretion disc is face-on to the observer. As such, the accretion disc's blue contribution to the system's \textit{near}-IR quiescent colours may be significant. Simple models of the nova ejecta have been constructed using the morphological modelling code \textit{XS5}, and the results support the above hypothesis. Precise spectral classification of this object has been exceptionally difficult owing to low signal-to-noise levels and high extinction, which has eliminated all evidence of any He/N or Fe~II emission within the spectra. We suggest two possibilities for the nature of V1721~Aql: that it is a U~Sco type RN with a sub-giant secondary or, less likely, that it is a highly energetic bright and fast classical nova with a main sequence secondary. Future monitoring of the object for possible RN episodes may be worthwhile, as would archival searches for previous outbursts.}

\maketitle
\section{Introduction and Observations} \label{sec:intro}
Classical novae (CNe) occur in interacting binary systems consisting of a white dwarf (WD) primary and a main sequence secondary star, which fills its Roche-lobe. They are a subclass of the cataclysmic variables (CVs). Hydrogen-rich material is transferred from the secondary and deposited onto the surface of the WD, usually via an accretion disc \citep[see][for recent review papers]{CNbook08}. A thermonuclear runaway eventually occurs within the accreted hydrogen-rich layer on the WD surface leading to a nova outburst \citep[see][]{Starrfieldpaper}. CNe tend to exhibit outburst amplitudes of approximately 10-20 magnitudes \citep{Shara} and eject $10^{-5}-10^{-4}$ M$_{\odot}$ of material at velocities between hundreds to a few thousand km $\mathrm{s^{-1}}$ \citep{Prialnik}, with outbursts occurring approximately once every $10^{4}-10^{5}$ years. If a nova system is seen to have more than one outburst, it is classed as a recurrent nova (RN). RNe undergo outbursts on a time-scale of decades up to $\sim$ 100 years and tend to have higher ejection velocities and lower ejected masses than CNe \citep{Anupama}. The secondary star in a RN system is often an evolved star such as a sub-giant or red giant \citep[see also review by][]{Bode10}. 

Novae can be grouped into classes depending on their speed of decline from maximum light \citep[``speed classes'',][]{PG} or the dominance of certain non-Balmer emission lines in their post-outburst spectrum \citep{Williams}. These emission lines are either those of Fe~II or He/N, yielding two spectral classes. The spectra of He/N novae tend to exhibit ``boxy'' structures and high expansion velocities, whereas Fe~II spectra are more Gaussian and have lower expansion velocities. Recurrent novae typically show features consistent with the He/N novae. 

In the Milky Way the distribution of novae shows a strong concentration towards the Galactic plane and the bulge. It has been shown by \citet{Della} that fast novae are more concentrated toward the Galactic plane (\textit{z} $<$ 100 pc) than slow novae, which are associated with the Galactic bulge extending up to 1 kpc. Fast novae are more luminous at peak than slow novae and possess more massive and luminous WDs. Work by \citet{Duerbeck} indicates that many of these bright novae, however, go undetected. The reasons for this are high interstellar extinction inherent in the Galactic plane, and that for many novae their speed of decline alone makes them difficult to detect \citep[see][for discussion]{Warner, Hounsell}. Even now, many fast highly extinguished novae are only ever detected and tracked by amateur astronomers.

Nova V1721~Aquilae ($\alpha=19^{h}06^{m}28^{s}\!\!.58,  \delta=+7^{\circ}06^{\prime}44^{\prime\prime}\!\!.3$; J2000) was discovered on 2008 September 22.5 UT by K. Itagkai. The outburst was confirmed on 2008 September 22.586 UT and reached a peak unfiltered magnitude of 14.0. Discovery of the nova was presented in \citet{Yamaokab} along with an initial spectral investigation. Post-outburst spectra were obtained on 2008 September 25.19 and 25.25 UT using the Steward Observatory Bok 2.29m telescope on Kitt Peak via the Boller $\&$ Chivens optical spectrograph (details of the instrumental set-up can be found in Section~\ref{subsec:spectra}). Initial analysis of the spectra revealed a broad triple-peaked H$\alpha$ emission profile with a full width half maximum (FWHM) of 6450 km s$^{-1}$, along with O~I 7773 \AA and O~I 8446 \AA structures \citep{Helton}. The ejecta velocities derived from the initial analysis of the spectra were very high indeed for a typical CN and there was initial suspicion that it may be a supernova (S. J. Smartt - private communication). Spectra also indicated that extinction towards the object was high and by comparison to other novae during similar evolutionary phases was estimated to be  $A_{V}\approx 9.3$. Hence, the distance to the nova was initially derived as 5 kpc, by assuming at maximum $M_{V} \approx-9$ \citep{Helton}. Due to the faintness of the source at maxima and its rapid decline, follow up spectroscopy of the object has not been possible. 


This paper aims to determine the nature of V1721~Aql through the examination of available photometric and spectroscopic data. Archival pre-outburst Two Micron All-Sky Survey\footnotemark[1] \citep[2MASS;][]{Skrutskie} photometry has enabled the determination of the spectral type of the secondary star within the system. Post-outburst photometric and spectroscopic data have been used to obtain the speed class of the nova, its extinction, the average ejection velocity of the system and potential spectral class. Section~\ref{sec:data} presents the analysis of the data; Section~\ref{sec:discussion} discusses the results found and the subsequent classification of the object.

\footnotetext[1]{The Two Micron All Sky Survey, is a joint project of the University of Massachusetts and the Infra-red Processing and Analysis Centre California Institute of Technology, funded by the National Aeronautics and Space Administration and the National Science Foundation.}

\section{Data Analysis}\label{sec:data}

\subsection{Distance determination}\label{subsec:distance}

\begin{figure}
\centering
\includegraphics[width=\columnwidth]{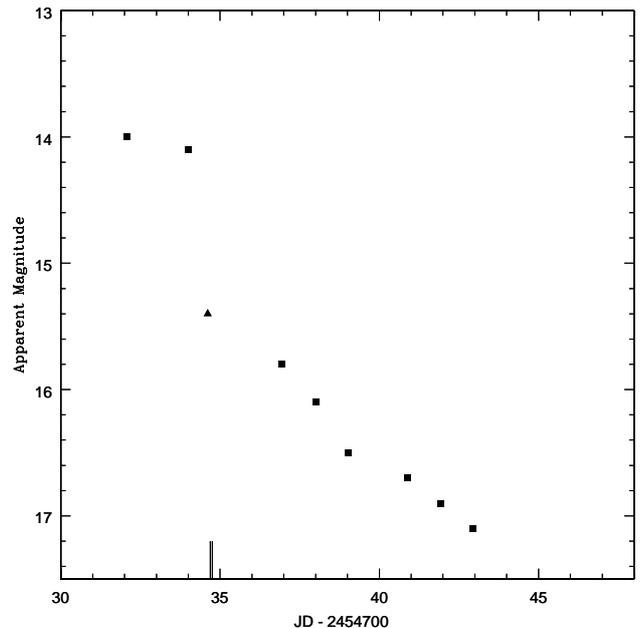}
\caption{Apparent magnitudes of Nova V1721~Aql as observed by K. Itagaki (squares - unfiltered) and R. King (triangles - Visual). These results are presented in VSNET\protect\footnotemark[2] and AAVSO\protect\footnotemark[3]. The two additional tick marks on the x-axis represent the dates on which the Blue (see Figure~\ref{figure2}) and Red (see Figure~\ref{figure3}) spectra were taken.}
\label{figure1}
\end{figure}

After outburst, V1721~Aql continued to be monitored photometrically until 2008 October 6 UT these results are reproduced in Figure~\ref{figure1}. These data indicate that $t_{2} \approx 6$ days for the nova, classifying it as very fast \citep{PG}. Using these data and the maximum magnitude-rate of decline relation \citep[MMRD,][]{Mclaughlin} with parameters from \citet{Downes}, we derive an absolute magnitude $M_{V} = -9.4 \pm 0.5$. 

The extinction towards V1721~Aql is thought to be extremely high, with as noted above, an estimate of $A_{V} \approx 9.3$ given in \citet{Helton}. This however, is based purely upon comparisons with other novae at a similar early evolutionary state. In order to obtain an independent extinction we made use of the \citet{Rowles} extinction maps which have a high spatial resolution and are able to detect a greater number of small-scale high extinction cores compared to other maps. These extinction maps are generated using 100 nearest neighbour stars and give an $A_{V} = 11.6\pm0.2$, much higher than the original estimate. Using the more accurate extinction from \citet{Rowles}, we derive a distance to V1721~Aql of $2.2\pm 0.6$ kpc.  

The Galactic coordinates of the object are l$^{II}$=41$^o$, b$^{II}$=-0.1$^o$. This indicates that nova V1721 Aql is located very close to the Galactic plane with \textit{z} = 2.5 pc, and is in a region of the sky in which it is typically very difficult to observe novae because of high extinction along the line of sight. 

\footnotetext[2]{Available from http://ooruri.kusastro.kyoto-u.ac.jp}
\footnotetext[3]{Available from http://www.aavso.org}

\subsection{Post-outburst Spectra}\label{subsec:spectra}

\begin{figure}
\centering
\includegraphics[width=\columnwidth]{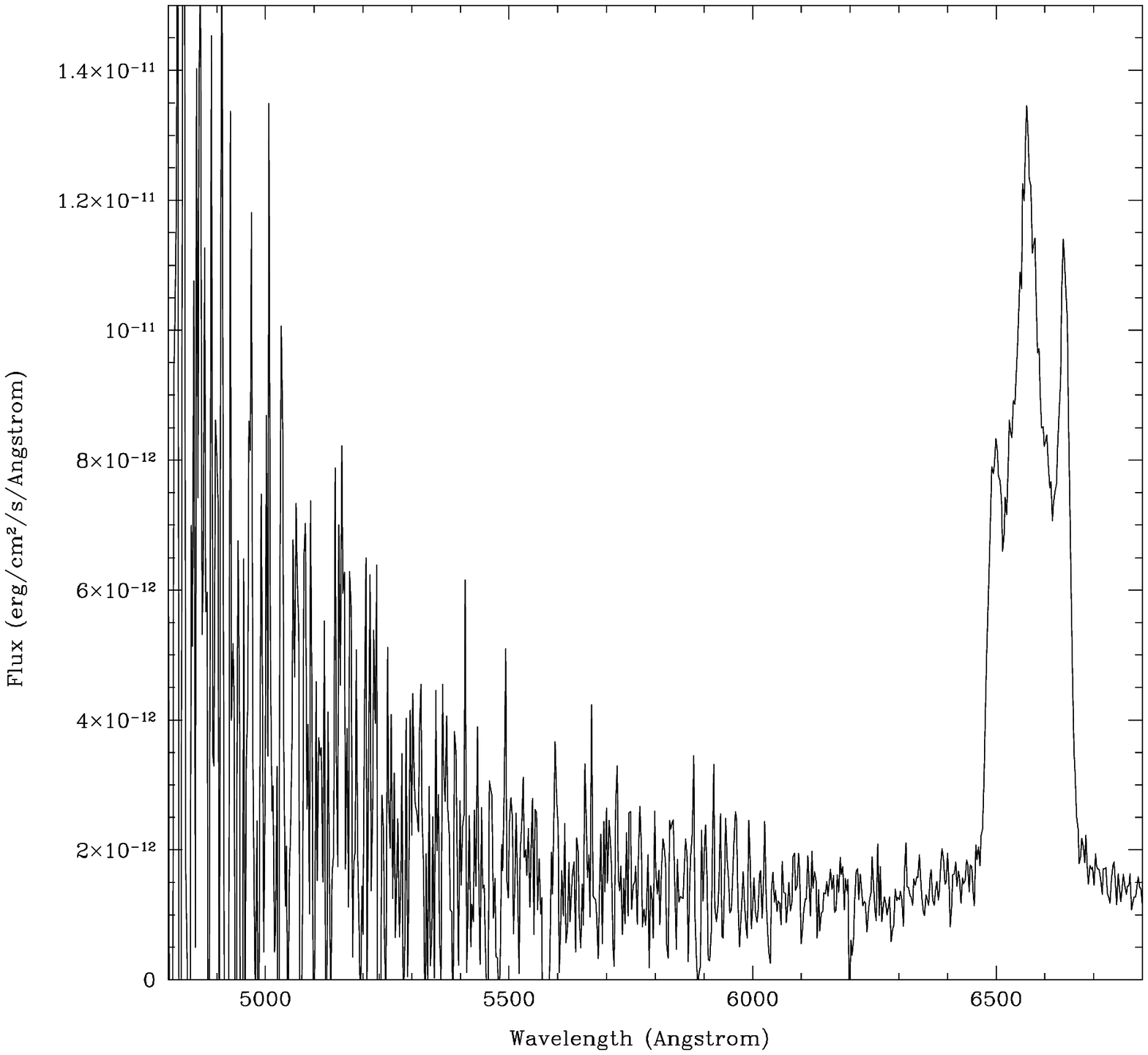}
\caption{Heliocentric velocity and extinction corrected ($A_{V}$ = 11.6, see Section~\ref{subsec:distance}) optical spectrum of V1721~Aql, taken on 2008 September 25.19 (2.69 days after discovery) with the Steward Observatory Bok 2.29m telescope.}
\label{figure2}
\end{figure}

\begin{figure}
\centering
\includegraphics[width=\columnwidth]{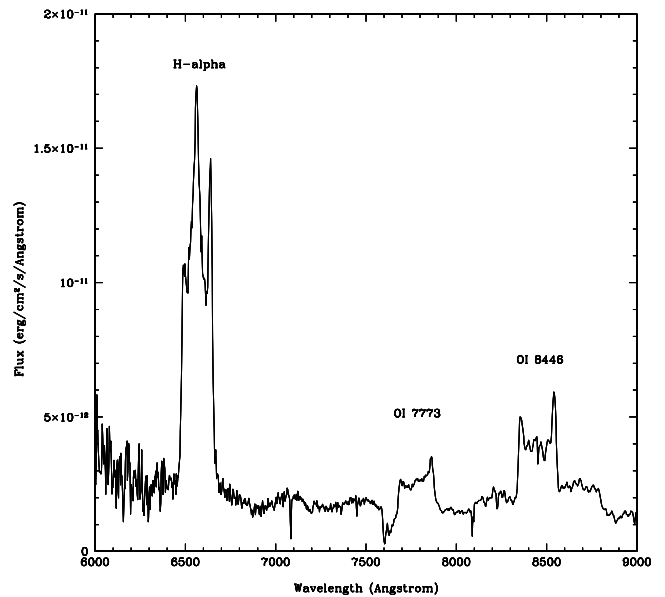}
\caption{Heliocentric velocity and extinction corrected ($A_{V}$ = 11.6, see Section~\ref{subsec:distance}) optical spectrum of V1721~Aql, taken on 2008 September 25.25 (2.75 days after discovery) with the Steward Observatory Bok 2.29m telescope.}
\label{figure3}
\end{figure}

Post-outburst spectra were obtained on 2008 September 25.19 and 25.25 UT using the Steward Observatory Bok 2.29m telescope on Kitt Peak with the Boller $\&$ Chivens optical spectrograph, and are presented in Figures~\ref{figure2} and \ref{figure3}. The ``Blue'' set-up utilised a 400 l mm$^{-1}$ 1$^{\mathrm{st}}$ order grating with a UV blocking filter to prevent order contamination below $\sim 3600$ \AA. The spectral coverage was from $\sim 3600$ to $\sim 6750$ \AA\ at a spectral resolution of roughly 2.8 \AA\ pixel$^{-1}$. The ``Red'' set-up was identical but with the grating centred near 7600 \AA\ providing coverage from $\sim 6000$ to $\sim 9250$ \AA\ and with a blocking filter effective below 4800 \AA. Flat fielding was performed using a continuum arc lamps. Red observations at wavelengths beyond $\sim 7700$ \AA\ are subject to fringing effects arising at the CCD that are unable to be corrected by flat fielding. The effect of this fringing on the data depends upon the target position on the sky and the target intensity. Wavelength calibration was performed using He-Ar-Ne calibration lamps at each target position. The spectroscopic standard Wolf 1346 was used for flux calibration. Spectra have also been corrected for heliocentric velocity and reddening ($A_V$ = 11.6). All data reduction was performed in IRAF\footnotemark[4] following standard optical data reduction procedures.

\footnotetext[4]{IRAF is distributed by the National Optical Astronomy Observatory, which is operated by the Association of Universities for Research in Astronomy (AURA) under cooperative agreement with the National Science Foundation.}

The Blue spectrum of V1721~Aql is presented in Figure~\ref{figure2}. It is important to note that this spectrum is devoid of detectable emission lines blue-wards of H$\alpha$, likely owing to the very high extinction. Because of the absence of H$\beta$ emission in the spectrum, a lower limit on the extinction is obtained using the Balmer decrement for Case B HI recombination and the observed intensity ratio of H$\alpha$ and H$\beta$. We note that this spectrum is taken early in the nova outburst and although the nova is very fast, conditions may not yet be those of Case B. From this we estimate a lower limit of A$_{V}$ $\geq$ 8. This value is consistent with both the above determinations of $A_{V}$ and helps to confirm that the extinction is indeed high. 

The Red spectrum of V1721~Aql is shown in Figure~\ref{figure3} and indicates the presence of a triple-peaked H$\alpha$ emission line along with emission structures corresponding to O~I 7773 \AA and O~I 8446 \AA. It is necessary to determine if the ``boxy'' structure around H$\alpha$ consists of purely H$\alpha$ or combined lines of H$\alpha$ $+$ [N~II] 6482, 6548, 6584, 6611 \AA. However, although [N~II] is expected in the spectra, this early in the outburst the [N~II] line strength is unlikely to be significant in comparison to H$\alpha$ and so an unlikely contributor to the boxy structure. An absence of [N~II] altogether may simply be owing to the fact that the nova has indeed been caught at maxima and the lines tend to develop a little later in the outburst. For example the fast, potentially U Sco like nova V2491 Cyg, showed evidence of [N~II] 4.62 days after peak magnitude, with these lines becoming more defined 32.7 to 108 days after peak \citep{Munari}. The lack of [N~II] 5755 \AA at shorter wavelengths also contradicts the idea of a strong N~II presence, although this line may have been missed due to a combination of the object's rapid evolution and high extinction. Additionally, the observed emission peaks at the blue and red edge of the H$\alpha$ profile in V1721 Aql are nearly symmetric and expected positions of potential [N~II] contaminants are not. 

\begin{figure}[ht]
\centering
\includegraphics[width=\columnwidth]{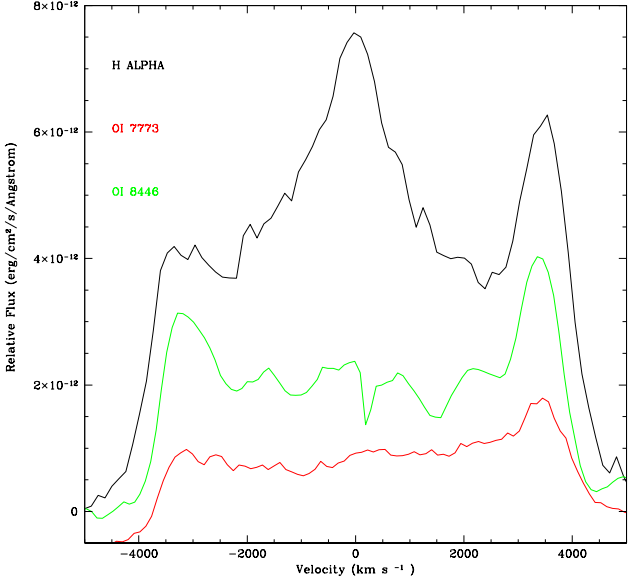}
\caption{Relative velocity diagram of the H$\alpha$, O~I 7773 \AA, and O~I 8446 \AA structures. Note that lines have been off-set on the y-axis for ease of comparison.} 
\label{figure4}
\end{figure}

A relative velocity diagram of the H$\alpha$, O~I 7773 \AA, and O~I 8446 \AA structures is given in Figure \ref{figure4}. This diagram indicates that the H$\alpha$ and O~I lines contain similar weak blue/strong red wing morphologies, however, the H$\alpha$ central peak is much more prominent and may arise in an emitting region distinct from the other components of the emission profile. The velocity shifts of the three components are also similar which supports the hypothesis that the H$\alpha$ structure consists of H$\alpha$ emission only. 

\begin{figure}[!ht]
\centering
\includegraphics[width=\columnwidth]{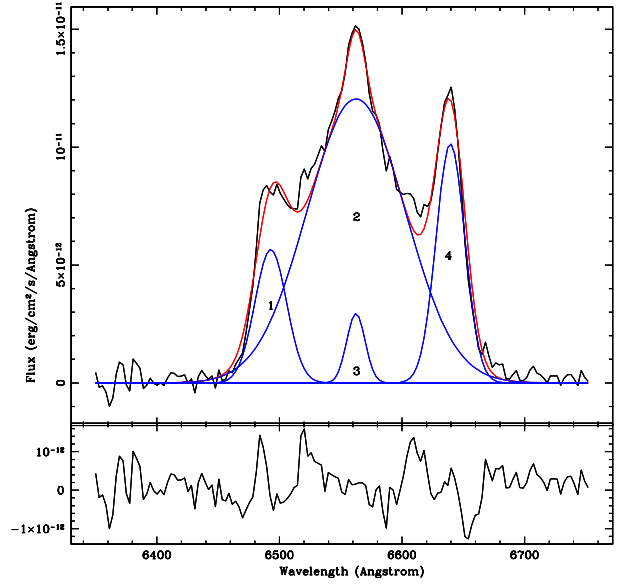}
\caption{Observed H$\alpha$ structure (black line) with the sum of \textit{Specfit} Gaussian components (red line). The blue lines represent separate Gaussian components. See Section~\ref{sec:discussion} for further discussion. The lower plot shows the residual to the fit. }
\label{figure5}
\end{figure}

\begin{figure}[!ht]
\centering
\includegraphics[width=\columnwidth]{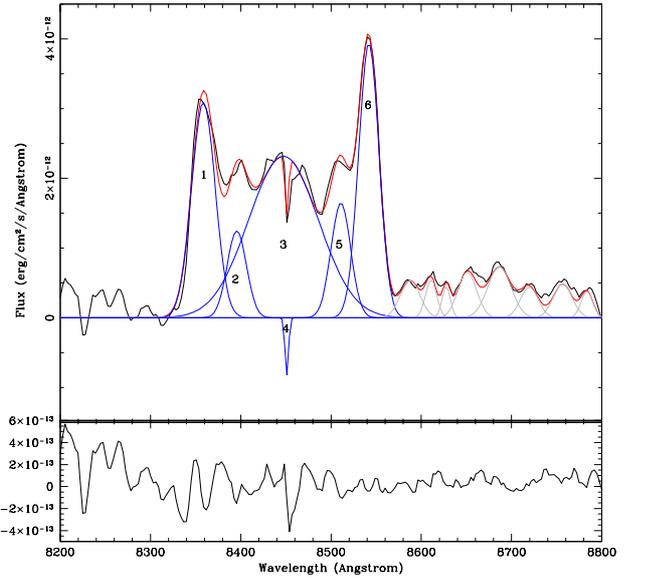}
\caption{Observed O~I 8446 \AA structure (black line) with the sum of  \textit{Specfit} Gaussian components (red line). The blue lines represent separate Gaussian components. Gaussian 4 represents a spectral artifact. Gaussians 2 and 5 represent components within the profile that are partially caused by fringing. Grey Gaussians represent fine structure caused by fringing effects. All fringing effects required fitting in order to produce the best overall match with observations. The lower plot shows the residual to the fit.}  
\label{figure6}
\end{figure}

In order to identify any potential emission lines that may be contaminating the H$\alpha$ structure a spectral fit of the region (using the Red spectrum, Figure~\ref{figure3}) was conducted using STSDAS's\footnotemark[5] \textit{Specfit}, the results of which are presented in Figure~\ref{figure5} and Table~\ref{table1}. It should be noted that the central H$\alpha$ peak was fit by two separate Gaussians with the second component added as a correction to the first in compensation for the oversimplification of the fit, a possible physical explanation for this is given in Section~\ref{sec:discussion}. The O~I 8446 \AA structure has also been fit with these results presented in Figure~\ref{figure6} and Table~\ref{table2}. It is evident that fringing occurs within the spectra at wavelengths $\gtrsim$ 8000 \AA. The effect of this fringing has been to contribute to components 2 and 5 of Figure~\ref{figure6} and to create fine structure short-ward of the O~I 8446 \AA profile. It was therefore necessary to fit these contaminating structures which would otherwise interfere with the results. Taking fringing effects into account the central structure of the O~I 8446 \AA line profile is most likely relatively flat. Although there are slight inherent differences in the strengths of the red/blue peaks in the profiles of the O~I 7773 and 8446 \AA features, the intrinsic shape before fringing effects of the 8446 \AA feature is likely very similar to the 7773 \AA feature, as illustrated in Figure~\ref{figure4}. The fitting of the O~I 7773 \AA profile has not been conducted as this structure has been severely truncated on the blue edge by an atmospheric absorption feature.

\footnotetext[5]{STSDAS is a product of the Space Telescope Science Institute, which is operated by AURA for NASA}

\begin{table*}
\caption{Wavelength, FWHM, and relative velocity of fitted components of the triple-peaked H$\alpha$ structure presented in Figure~\ref{figure5}.}            
\label{table1}      
\centering                 
\begin{tabular}{c c c c}        
\hline\hline                 
Gaussian & Wavelength (\AA) & FWHM (km s$^{-1}$) & Relative velocity (km s$^{-1}$)\\   
\hline                     
1 & 6493$\pm$3 & 1400$\pm$100 & -3200$\pm$100\\
2 & 6563$\pm$1 & 4300$\pm$100 & -10$\pm$50\\
3 & 6563$\pm$1 & 800$\pm$100 & -10$\pm$50\\
4 & 6639.6$\pm$0.4 & 1260$\pm$50 & 3510$\pm$20\\
\hline                                
\end{tabular}
\end{table*}

\begin{table*}
\caption{Wavelength, FWHM, and relative velocity of primary fitted components of the triple peaked O~I 8446 \AA structure presented in Figure~\ref{figure6}.}            
\label{table2}      
\centering                      
\begin{tabular}{c c c c}        
\hline\hline                 
Gaussian & Wavelength (\AA) & FWHM (km s$^{-1}$) & Relative velocity (km s$^{-1}$)\\   
\hline                     
1 & 8358.7$\pm$0.5 & 1110$\pm$40 & -3100$\pm$20\\
3 & 8447$\pm$1 & 3200$\pm$200 & 40$\pm$50\\
6 & 8541.9$\pm$0.4 & 930$\pm$30 & 3410$\pm$10\\ 
\hline                                
\end{tabular}
\end{table*}

We find that no other spectral lines expected to be found in novae match the wavelengths presented within Tables~\ref{table1} and \ref{table2}. Given this, and that the blue and red wings of the H$\alpha$, O~I 7773, and O~I 8446 profiles are similar, we conclude that these structures consist of H$\alpha$ and O~I only. Combining the relative velocities of H$\alpha$ Gaussians 1 and 4 gives a mean expansion velocity of $V_{exp} = 3400\pm200 km s^{-1}$. The structure of each line profile may also tell us something about the nova ejecta geometry (see Section~\ref{subsec:secondary} for further discussion).

On examination of both Blue and Red spectra, no evidence of Fe~II/[Fe~II] was found. This could be caused by the faintness of the spectra, noting the high extinction to the object, and hence the high noise level. There may be a some evidence of He~I 7001 \AA and N~I 8680, 8703, 8711 \AA emission. However, due again to noise within the spectra and fringing effects at these longer wavelengths, it is difficult to calculate their significance. Exact spectral classification of the object according to the \citet{Williams} system therefore remains elusive.

\subsection{Pre-outburst identification}\label{subsec:preout}
Pre-outburst images of a source at the location of V1721~Aql are found within the 2MASS catalogue\footnotemark[6], with \textit{near}-IR co-ordinates given as $\alpha=19^{h}06^{m}28^{s}\!\!.60, \delta=+7^{\circ}06^{\prime}44^{\prime\prime}\!\!.46$; J2000. Observed 2MASS apparent magnitudes and colours of the \textit{near}-IR source located at the position of the nova can be found in Table~\ref{table3}. This table also contains de-reddened colours using the extinction value $A_{V}= 11.6\pm 0.2$. 

\begin{table*}
\caption{2MASS apparent and absolute magnitudes of V1721~Aql progenitor (candidate), colours, and de-reddened colours of the \textit{near}-IR source located at the position of the nova. The extinction towards the nova has been taken as A$_{V}$ = 11.6, and its distance as 2.2 kpc.}            
\label{table3}      
\centering                       
\begin{tabular}{c c c | c c c }        
\hline\hline                 
Filter & Apparent magnitude & Absolute Magnitude & Colour & Value & De-reddened Value\\
\hline                     
   
  \textit{J} & $16.6\pm0.2$ & $1.8\pm0.6$ & \textit{J-K$_{s}$} & $2.0\pm0.2$ & $0.1\pm0.2$\\    
  \textit{H} & $15.5\pm0.1$ & $1.7\pm0.6$ & \textit{J-H} & $1.2\pm0.2$ & $0.0\pm0.2$\\ 
  \textit{K$_{s}$} & $14.7\pm0.1$ & $1.7\pm0.6$ & \textit{H-K$_{s}$} & $0.8\pm0.2$ & $0.1\pm0.2$\\
   
\hline                           
\end{tabular}
\end{table*}

\begin{figure*}
\centering
\subfigure[]{
\includegraphics[width=0.45\textwidth]{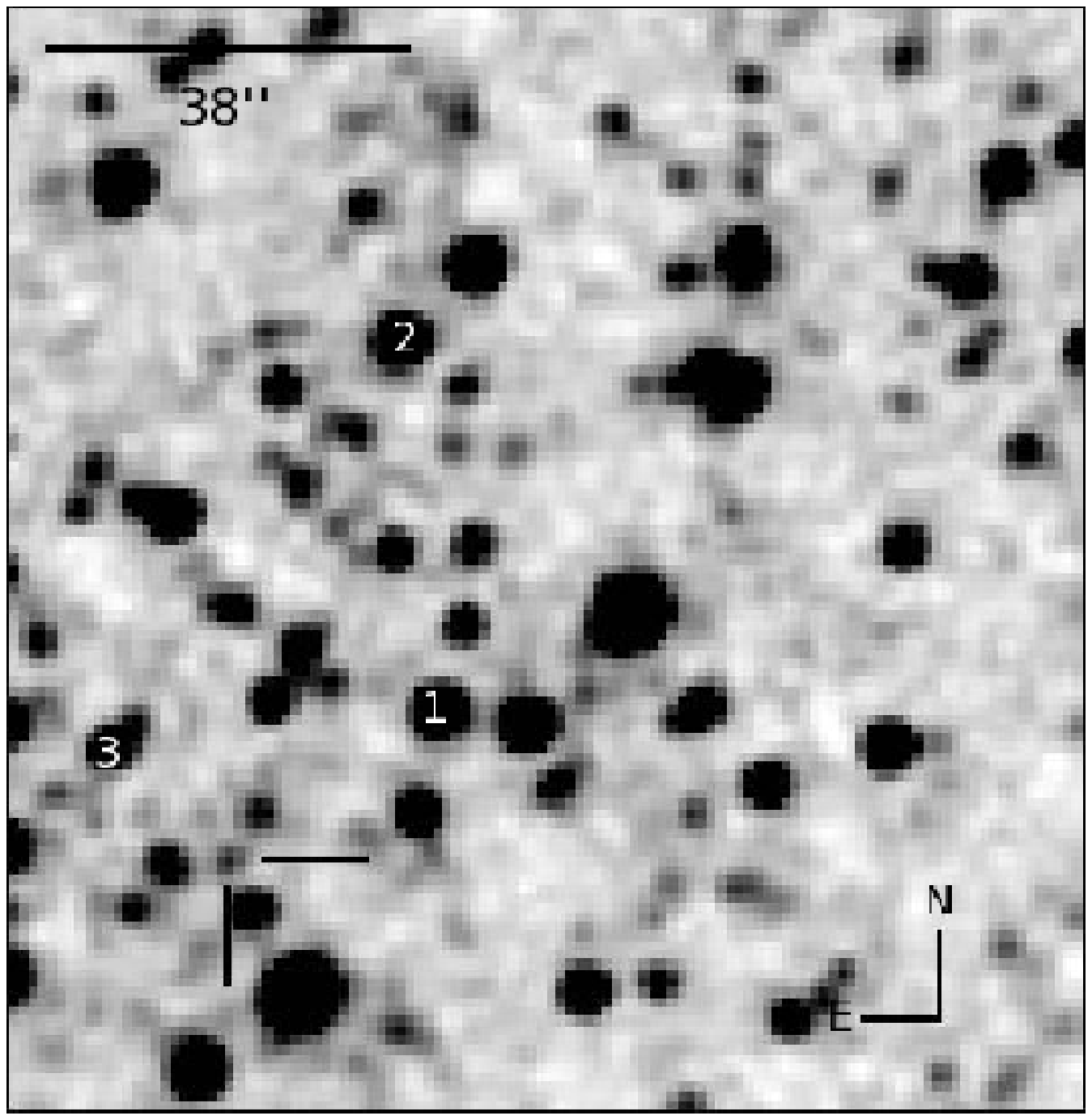}
\label{figure7a}
}
\subfigure[]{
\includegraphics[width=0.45\textwidth]{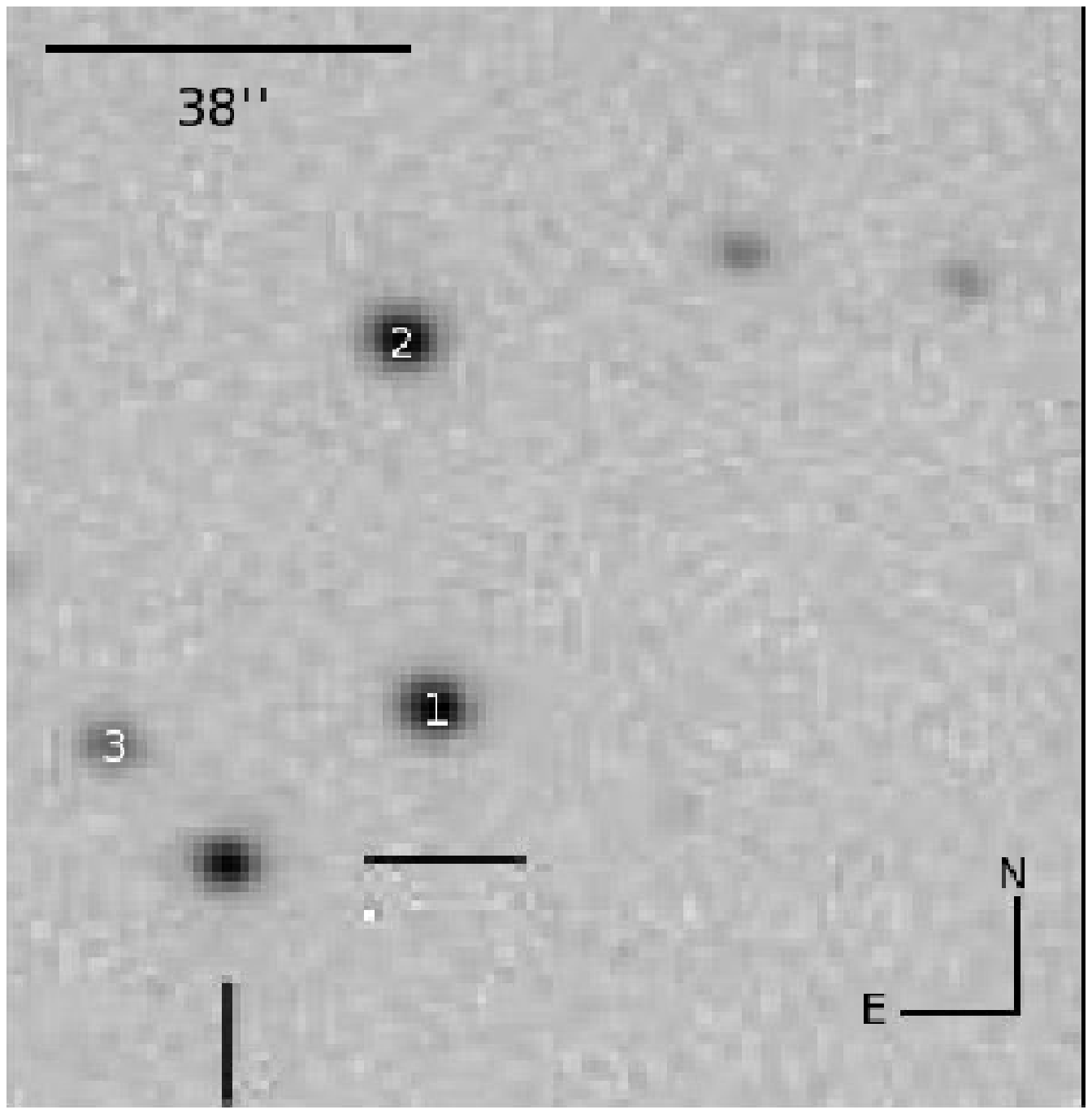}
\label{figure7b}
}
\caption{(a) 2MASS \textit{K$_{s}$} band pre-outburst image. The nova is found within the centre of the black markers. (b) Unfiltered discovery image taken by K. Itagka on the 22nd of September at the Itagaki Astronomical Observatory. The nova is found within the centre of the black markers. }
\label{figure7}
\end{figure*}

The V1721~Aql discovery image\footnotemark[7] and 2MASS \textit{K$_{s}$} image were aligned and compared via IRAF packages. Based on the stellar density within the 2MASS \textit{K$_{S}$} pre-outburst image the probability of a chance alignment at least as close as that found between the nova and the 2MASS object is less than 1$\%$. The archival 2MASS \textit{K$_{s}$} band image is presented in Figure~\ref{figure7}(a) and the discovery image presented in Figure~\ref{figure7}(b).

\footnotetext[6]{Available from http://irsa.ipac.caltech.edu/applications/Gator/}
\footnotetext[7]{Available from http://www.astroalert.su/2008/09/24/nova-aql-2008}

\subsection{The Nature of the Secondary}\label{subsec:secondary}
There are several factors that contribute to the observed \textit{near}-IR colours of a nova system in quiescence, (i) the spectral type  of the secondary and evolutionary phase, (ii) the rate of mass transfer $\dot{M}$, (iii) the extinction $A_{V}$, (iv) the accretion disc and its inclination \textit{i}, and (v) the mass of the primary. In CNe one would expect that the effect of the emission from the WD on the \textit{near}-IR colours to be negligible, and the accretion disc to only provide a significant contribution to the emission when \textit{i} $\lesssim 30^{\circ}$, where an angle of $\textit{i} = 90^{\circ}$ is defined as an edge-on accretion disc \citep{Weight}. The location of a quiescent nova on a \textit{near}-IR two-colour diagram (\textit{$H-K_{s}, J-H$}) is therefore an important determinant of the nature of the secondary star in the system. 

\begin{figure}
\centering
\includegraphics[width=\columnwidth]{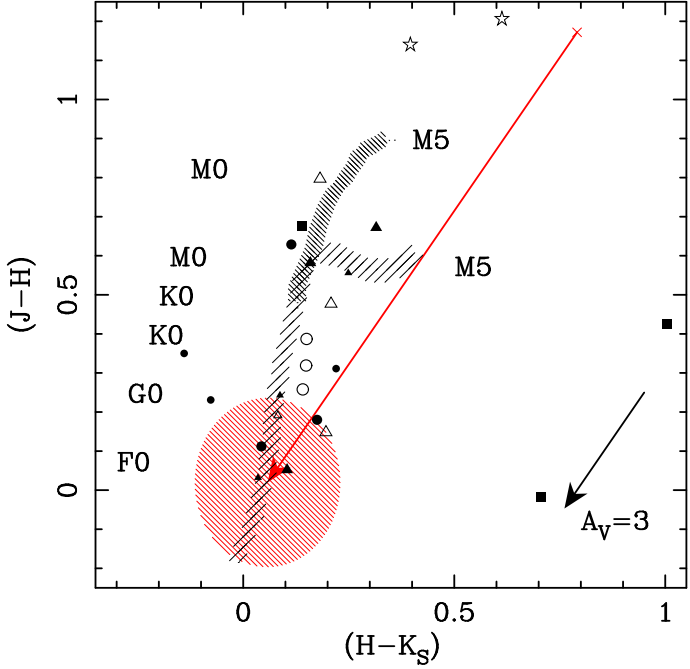}
\caption{\textit{Near}-IR colour-colour diagram of quiescent classical nova systems reproduced from Figures 4 $\&$ 7 in \citet{Hoard} using Table 1 of their data. The figure is adjusted to include the quiescent 2MASS colours of the nova V1721~Aql system. The light cross-hatched area represents \textit{near}-IR colours of main sequence stars with the denser cross-hatched area representing the giant branch \citep[see references within][]{Hoard}. The black points show individual nova systems and are coded according to the time since outburst, $\tau$, as follows; filled squares: $\tau < 25$ years; filled triangles: $\tau = 25-50$ years; open triangles: $\tau = 50-75$ years: filled circles: $\tau = 75-100$ years; open circles: $\tau > 100$ years. The star-shaped points are the recurrent nova systems. The nova systems presented here have not been corrected for extinction as in most cases the reddening is not accurately known, but it is assumed to be small to negligible in the \textit{near}-IR for most Galactic novae systems. The large points for individual nova systems have 1$\sigma$ uncertainties of $\leq 0.1$ magnitudes, smaller points have 1$\sigma$ uncertainties of $>$ 0.1 magnitudes. The red cross represents the observed quiescent \textit{near}-IR colours of the V1721~Aql nova system. The red line indicates the system's reddening vector with the arrowhead indicating its \textit{near}-IR colours once corrected for an extinction of $A_{V}$ = 11.6. The region enclosed by the red cross-hatching indicates all colours the nova system could possess within the error circle of $A_{V} = 11.6 \pm 0.2$. A de-reddening vector corresponding to $A_{V}$ = 3 is also shown.}
\label{figure8}
\end{figure}

The \textit{near}-IR apparent colours of the V1721~Aql nova system in quiescence are shown in Figure~\ref{figure8}. The system's colours occupy a region which contains the RN V745~Sco, which has a giant secondary with an M5+ III spectral type, and the suspected recurrent V1172~Sgr \citep{Weight}, which is also thought to contain a giant secondary. The extinction of these novae however, is much lower \citep[for V745~Sco $A_{V} = 3.1 \pm 0.6$;][]{Schaefer} than that of V1721~Aql. Nova V1721 Aql's occupancy of this region is merely coincidental and does not indicate that it is a RN-like system. 
Nova Aql's de-reddening vector is indicated with a red line, the arrow head on this line represents an extinction value of $A_{V} = 11.6$, the surrounding red region represents the error circle of the corrected colours. The de-reddened quiescent \textit{near}-IR colours of the V1721~Aql nova system lie within a region occupied by many quiescent CNe. Assuming that the \textit{near}-IR emission of the nova system is dominated by the secondary, Figure~\ref{figure8} indicates that its spectral type is that of a late F-G (possibly K) main sequence star. However, the inclination of the accretion disc must be taken into account and if it is less than 30$^{\circ}$ (approaching face-on to the observer) then its contribution to the \textit{near}-IR colours would be to cause a significant blue-wards offset. 

The line profiles observed in the nova spectra, and resultant high velocities, would suggest that the inclination of the disc within the binary system is low (face-on). The blue and red peaks seen within the H$\alpha$ and O~I structures would therefore be the result of material ejected along the poles towards and away from the observer \citep[observations and shaping models predict that the minor axis of a remnant lies in the disc plane;][]{Slavin, Porter}. This inclination however, would mean that the contribution by the disc to the \textit{near}-IR colours is significant. Taking this blue contribution into account shifts the \textit{near}-IR system colours along the main sequence and into the sub-giant region. Based on the speed and luminosity of the nova, the object may therefore be thought of as a U~Sco type RN system. Comparisons at quiescence between the absolute \textit{J} band magnitudes and \textit{H-K$_{S}$} colours of V1721~Aql (see Table~\ref{table3}), U~Sco (M$_{\textit{J}} = 1.3\pm0.4, H-K_{S} = 0.0\pm0.1$\footnotemark[8]), and V2491~Cyg ($M_{\textit{J}} = 1.0 \pm 0.3$, Darnley et al., submitted 2011, a suspected recurrent nova belonging to the U~Sco class) support this argument as they all possess similar absolute \textit{J} magnitudes and occupy the same region of space in an equivalent colour magnitude diagram around the sub-giant branch. The probability of a red giant as the secondary can also be ruled out as the \textit{J} band absolute magnitude of the system would have to be approximately five magnitudes brighter.

\footnotetext[8]{Photometry taken from \citep{Hanes}, distance and extinction taken from \citep{Schaefer}}

\begin{figure}
\begin{center}
\subfigure[]{\includegraphics[width=0.45\columnwidth]{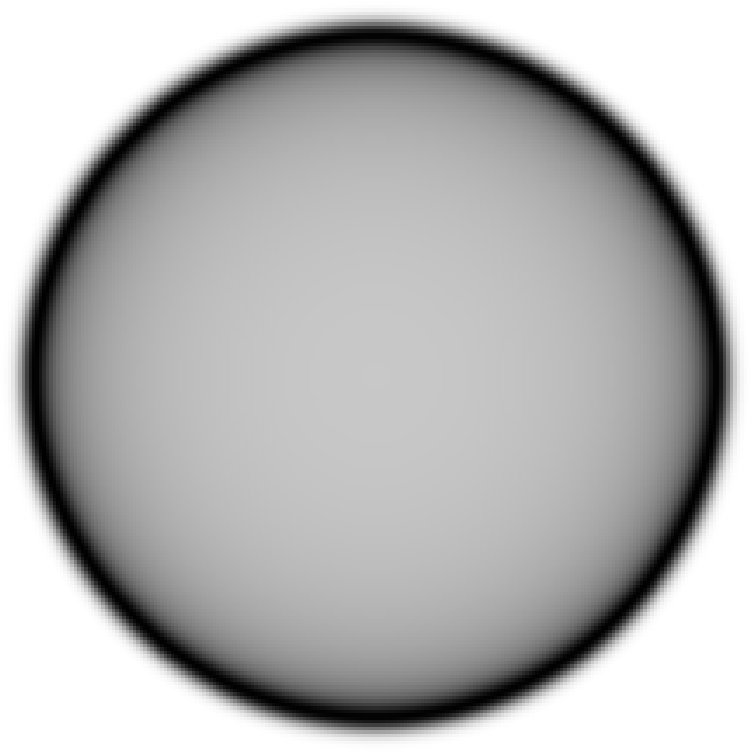}}
\subfigure[]{\includegraphics[width=0.45\columnwidth]{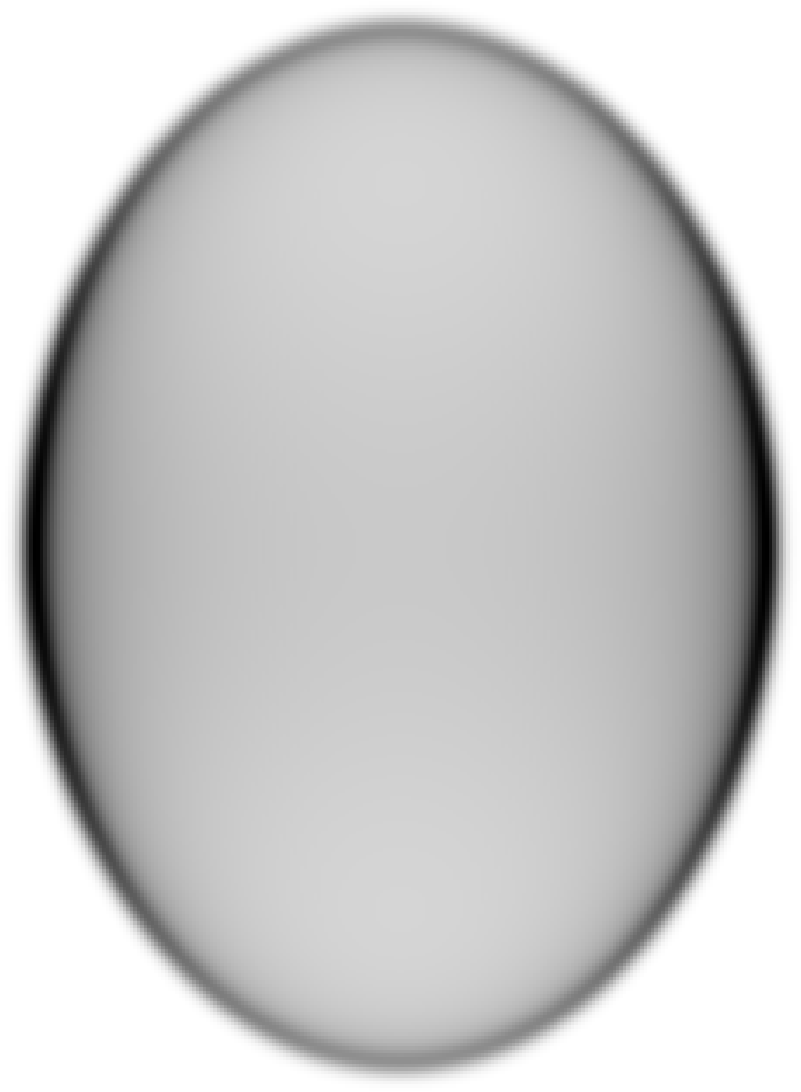}}\\
\subfigure[]{\includegraphics[width=\columnwidth]{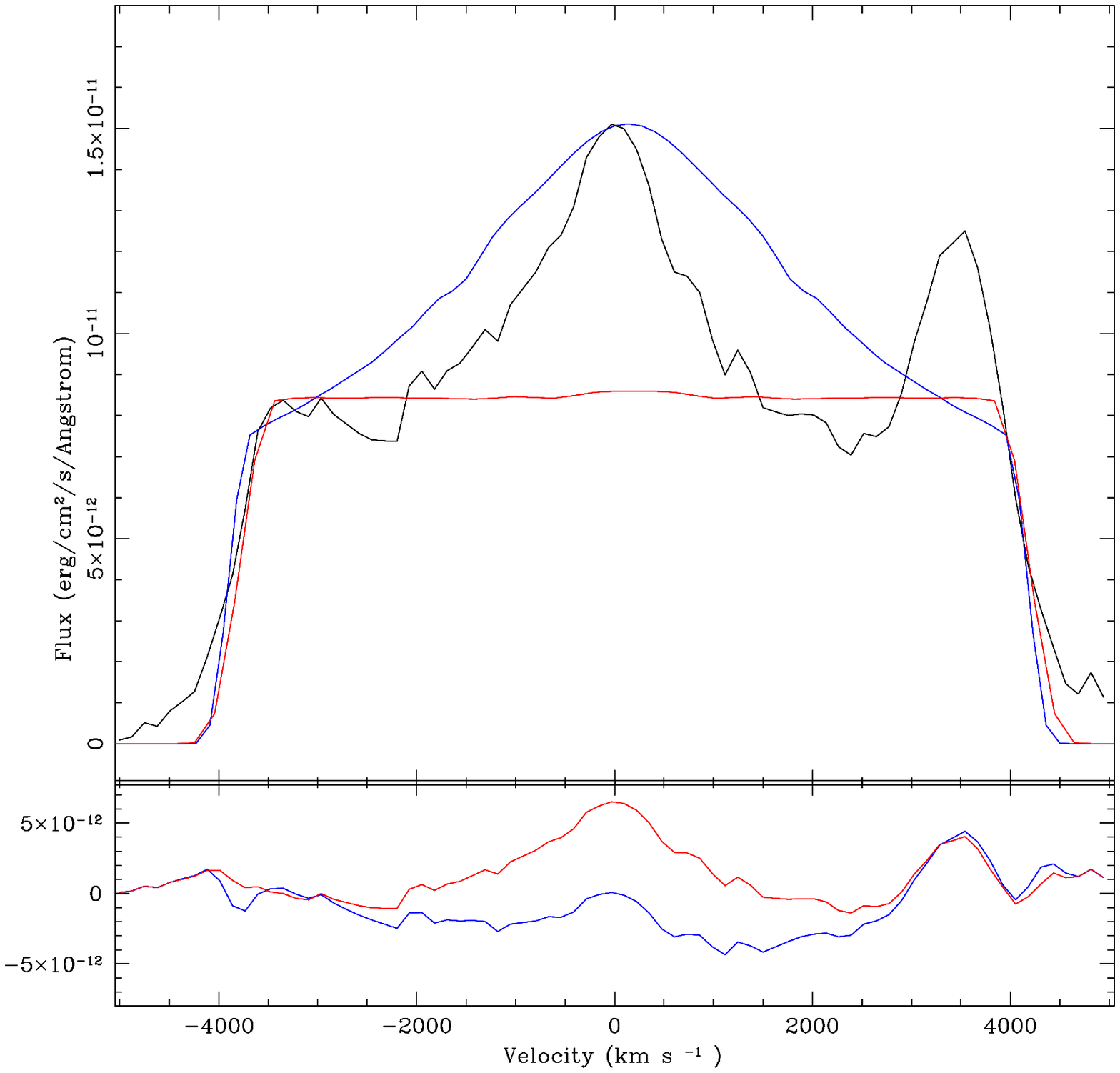}}
\end{center}
\caption{(a) Plan view of modelled nova ejecta with an axis ratio of 1, the shell is smooth with uniform emission. (b) Plan view of modelled nova ejecta with an axis ratio of 1.4, inclination of the system is such that the accretion disc is face-on to the observer, the shell is smooth with uniform emission . (c) Relative velocity diagram of the observed H$\alpha$ (black) structure and the two modelled systems created in \textit{XS5}. The red line represents the system with an axis ratio of 1, the blue line represents the system with an axis ratio of 1.4. The difference between the modelled and observed line profiles is given in the lower part of the diagram.}
\label{figure9}
\end{figure}

\begin{figure} 
 \begin{center}
\subfigure[]{\includegraphics[width=0.45\columnwidth]{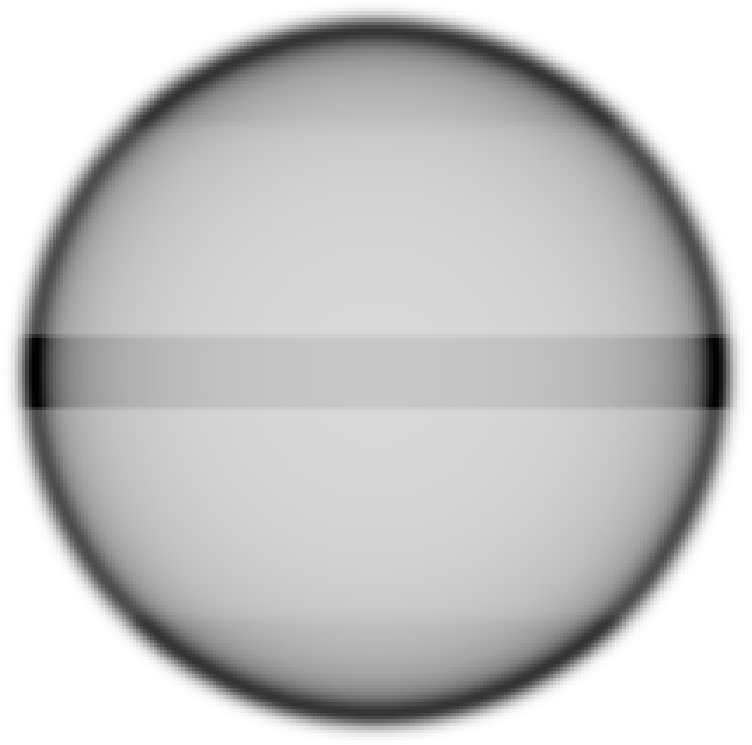}}
\subfigure[]{\includegraphics[width=0.45\columnwidth]{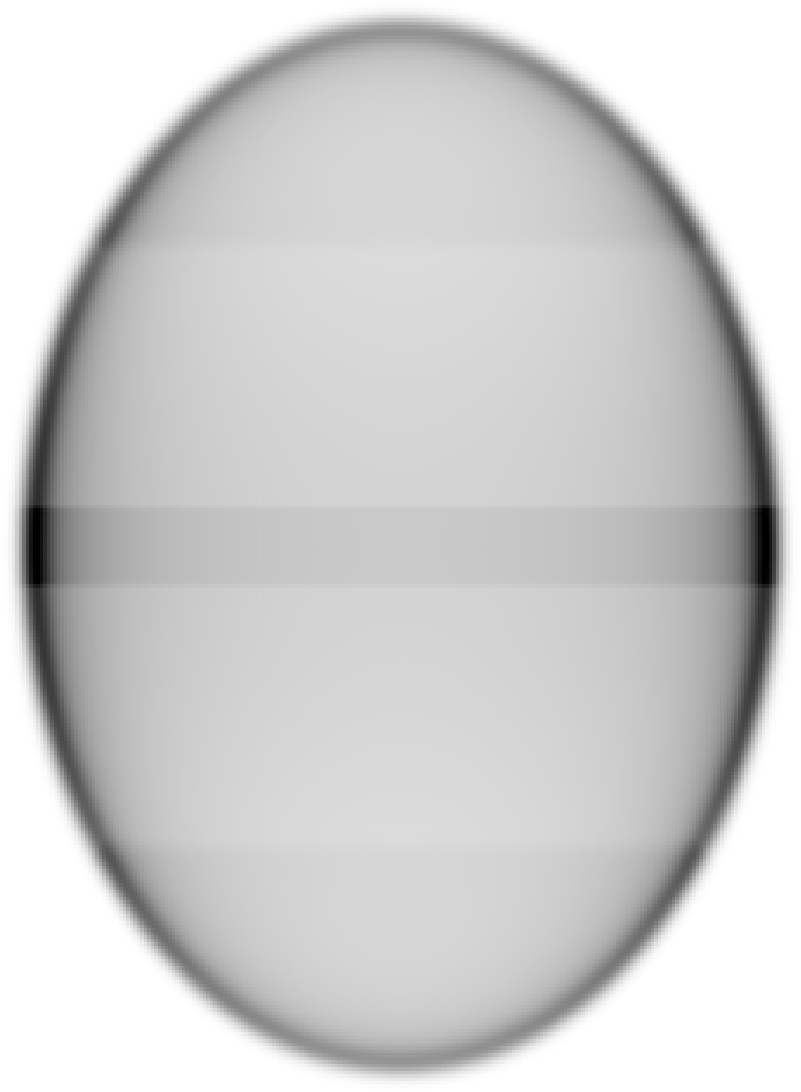}}\\
\subfigure[]{\includegraphics[width=\columnwidth]{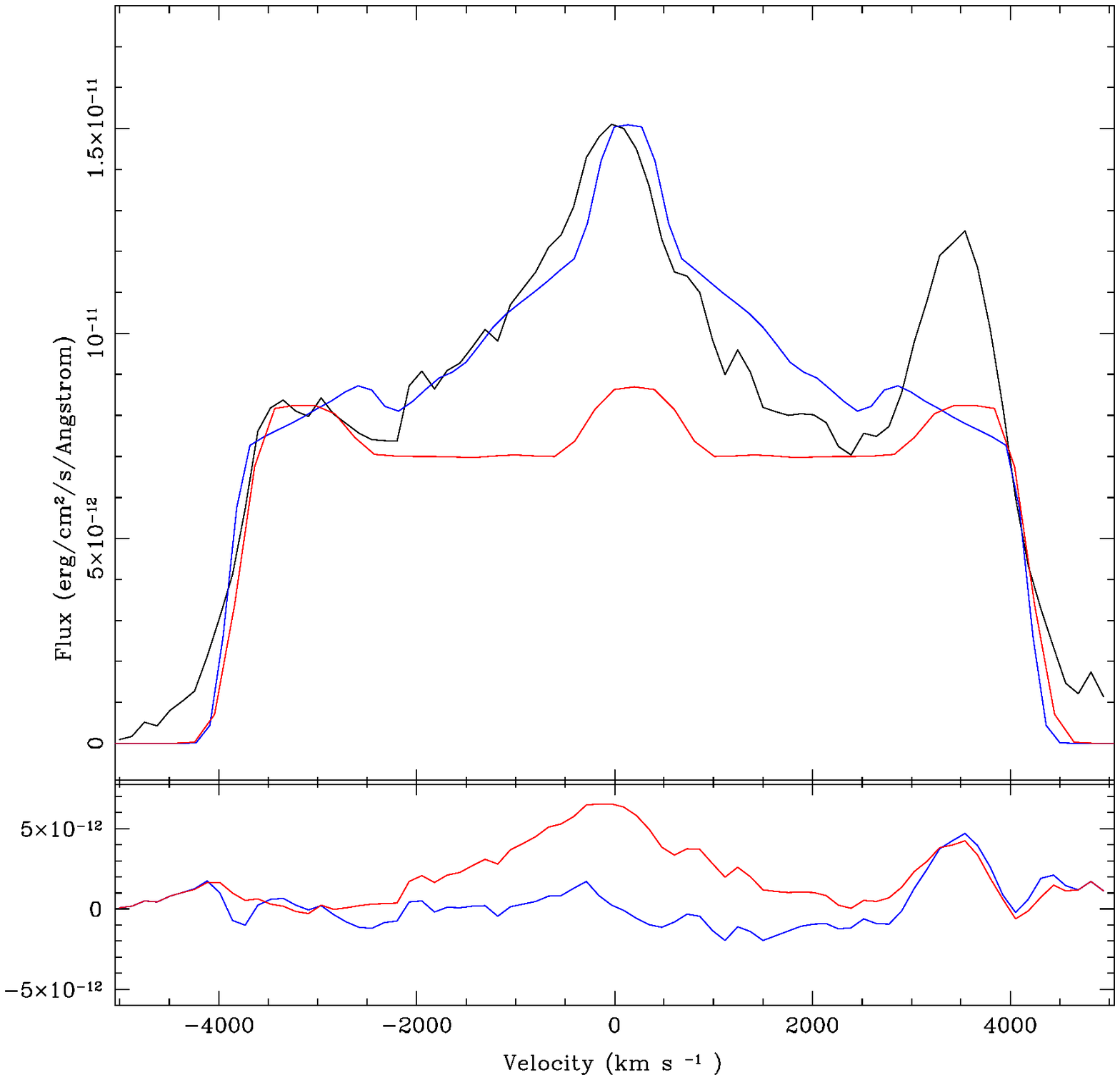}}
\end{center}
\caption{(a) Plan view of modelled nova ejecta with an axis ratio of 1, inclination of the system is such that the accretion disc is face-on to the observer, the shell is smooth with a slight emission enhancement within the equatorial region. (b) Plan view of modelled nova ejecta with an axis ratio of 1.4, inclination of the system is such that the accretion disc is face-on to the observer, the shell is smooth with a slight emission enhancement within the equatorial region. (c) Relative velocity diagram of the observed H$\alpha$ (black) structure and the two modelled systems created in \textit{XS5}. The red line represents the system with an axis ratio of 1, the blue line represents the system with an axis ratio of 1.4. The difference between the modelled and observed line profiles is given in the lower part of the diagram.}
\label{figure10}
\end{figure}

Given the speed of decline of the nova, work by \citet{Slavin} would suggest that the axis ratio (ratio of semi-major to semi-minor axis) of Nova V1721~Aql's ejected shell is low (\textbf{$\approx$1}). The nova ejecta may therefore be modelled by an approximately spherical shell with discreet randomly distributed knots of brighter emission. We attempted to model such a nova system by calculating the expected emission line profiles from models of the ejecta distribution and comparing them to observed profiles, specifically H$\alpha$. In order to do this we used \textit{XS5}, a morphological and kinematical modelling programme \citep{Dan} for producing 3D representations of astrophysical shells, synthetic images, and spectra. This program allows the user to generate a geometrical shape, such as an ellipsoid or an hourglass, which can be rotated and inclined. By adjusting additional parameters, such as the major and minor axis lengths, the FWHM of line profiles from the shell, the polar axis emission gradient, and the expansion velocity, the output emission line profile can be altered until a match with observations is achieved. Models of the nova ejecta with axis ratios between 1.0 and 2.0 (at 0.1 increments) were created. The results of this program are presented in Figures~\ref{figure9} and \ref{figure10}. Figure~\ref{figure9}(c) presents two modelled spectra compared to the observed H$\alpha$ structure. The red spectrum is that of a spherical shell, axis ratio of 1 and the blue spectrum is of an ellipsoidal-like shell with an axis ratio of 1.4. Both shells are smooth with uniform emission, and the inclination of the system is such that the central accretion disc is face-on. Figure \ref{figure10}(c) illustrates the results from the same two structures with the same inclination, but this time there is a slight emission enhancement in the equatorial region. From these modelled spectra it would seem that an ellipsoidal-like morphology may actually be more suited to the V1721~Aql ejecta, however we have far too little information to make a strong argument for this. We note that this higher axis ratio is contradictory to expectations in \citet{Slavin}. However, recent work on the 2010 outburst of U~Sco by \citet{Drake} has indicated that nova ejecta can be significantly shaped by circumbinary gas and/or a high accretion disc gas density. We have also been unable to reproduce the stronger red peak of the H$\alpha$ emission line profile. This could possibly be due to clumps in the ejecta, but more detailed data and modelling are needed to explore this further.

\section{Discussion and Conclusions}\label{sec:discussion}

The results presented in this paper indicate that V1721~Aql is a very fast nova (t$_{2} \sim$ 6 days) and very luminous ($M_{V} = -9.4 \pm 0.5$). The extinction of the object is high, $A_{V} = 11.6 \pm 0.2$ as the nova is very close to the Galactic plane. Based on the value of $A_{V}$, the distance to the nova is estimated to be $2.2 \pm 0.6$ kpc

Pre-outburst \textit{near}-IR colours of the nova have been compared to other novae in quiescence (all post-outburst) and the \textit{near}-IR colours of main sequence and giant stars. The results indicate that, when de-reddened, the nova occupies a region of the colour-colour phase-space in which most CNe are found and appears to have a late (F-M) main sequence secondary or a sub-giant. However, we cannot rule out the possibility of V1721~Aql being a RN, only that it does not appear to contain a giant secondary and can therefore not belong to the RS~Oph or T~CrB class of recurrent. The U~Sco class of RNe however, consists of an evolved main sequence or sub-giant secondary, much like CNe, and like V1721~Aql, the novae are very fast. Similarities in absolute \textit{J} band magnitudes at quiescence between V1721~Aql, U~Sco, and V2491~Cyg (suspected U~Sco member) also indicate that this object may be a U~Sco type RN.

Post-outburst spectra of the V1721 Aql revealed boxy structures around H$\alpha$, O~I 7773, and O~I 8446 \AA. We note that similar complex H$\alpha$ profiles have been observed in other fast novae such as the 1999 outburst of U Sco \citep{Iijima}, and the 2009 nova V2672 Oph \citep{Munari2}, a suspected U Sco type object. Examination indicates that the features in V1721 Aql are not contaminated significantly by other emission lines. The structure of the line emission would suggest that material is being ejected from the poles of the nova shell moving towards and away from the observer, leading to the blue and red wing emission seen. This would indicate that the disc of the binary is face-on, an argument which is also supported when considering the high observed velocity of the system. If the accretion disc were edge-on, concealed velocities would be greater than those observed, this is unlikely. A face-on accretion disc is also more likely when considering the physical reasons for Gaussian 3 of Figure~\ref{figure3}. The Gaussian may represent a narrow core of H$\alpha$ emission, in which case we are seeing H recombination emission both from the expanding ejecta, which gives rise to the broad overall emission, and emission from a reestablished accretion disc, or possibly even a disc that was never completely disrupted.

With a face-on accretion disc the hotter inner region of the disc is exposed possibly giving a significant blue contribution to the \textit{near}-IR colours of the nova and thus seriously affecting previous spectral classification of the secondary. The speed of decline of the nova also suggest that the nova shell itself has a low axis ratio so that it is almost spherical. Basic models generated by \textit{XS5} to reproduce the H$\alpha$ line profile, however, produce a best fit when using an ellipsoidal-like shell with an axis ratio 1.4. The departure from a spherical shell is most likely due to the nature of the explosion environment \citep[circumbinary material and/or high density disc gas;][]{Drake}. The model shell produced is smooth with a slight emission enhancement within the equatorial region, implying a face-on central accretion disc.



Relative velocity shifts found via spectral fitting of the H$\alpha$ and O~I emission are comparable to those presented in \cite{Helton} and we estimate an ejecta expansion velocity of $V_{exp} = 3400\pm$ 200 km s$^{-1}$ along the line of sight. This $V_{exp}$ is, perhaps, more in consistent with that of a fast classical nova system rather than fast recurrent novae, which tend to have slightly higher expansion velocities of $V_{exp} \gtrsim$ 4000 km s$^{-1}$ \citep{Anupama}.

There is no evidence of emission blue-wards of the H$\alpha$ structure. This is likely due to the high extinction towards the nova. An alternative explanation is that the nova is of the Fe~II class and at this stage the shell is still optically thick.
However, no dominant lines of Fe~II or [Fe~II] are present within the spectra. These lines may not have developed yet or may have been lost within the noise of the spectra. Therefore, an Fe~II classification can not be ruled out. Another complication for this hypothesis is that Fe~II novae tend to be slower than V1721~Aql.

The spectra show no conclusive evidence of He and N emission. This may also be due to the low signal to noise level within the spectra and the high extinction.   
No absorption features are seen in the spectra. This is unusual as we would expect to see absorption lines within the optically thick expansion stage.

In conclusion the precise nova sub-class of this object remains elusive, and the results of this work suggest two possibilities. The first is that this is a highly energetic luminous and fast classical nova with a main sequence secondary of spectral type F-M, and that any Fe~II lines that may have been observable within the nova spectra have simply been extinguished. The second possibility is that this is a U~Sco type RN and that evidence of He/N within the spectra is lost due to the high extinction. The latter scenario may prove itself within the next few decades and therefore this object is one that merits continued monitoring for future outbursts.

\begin{acknowledgements}
This publication makes use of data products from the Two Micron All Sky Survey, which is a joint project of the University of Massachusetts and the Infra-red Processing and Analysis Center/California Institute of Technology, funded by the National Aeronautics and Space Administration and the National Science Foundation.
We would like to thank Stephen Smartt and Rubina Kotak of Queen's University Belfast for pointing out this very interesting nova and providing us with further information. We would also like to thank an anonymous referee for detailed and thoughtful comments that have helped improve the paper.
R. Hounsell is supported by a PhD studentship from the Science and Technology Facilities Council of the UK.

\end{acknowledgements}

\bibliographystyle{aa} 

\begin{thebibliography}{29}
\expandafter\ifx\csname natexlab\endcsname\relax\def\natexlab#1{#1}\fi

\bibitem[{{Anupama}(2008)}]{Anupama}
{Anupama}, G.~C. 2008, in Astronomical Society of the Pacific Conference
  Series, Vol. 401, RS Ophiuchi (2006) and the Recurrent Nova Phenomenon, ed.
  {A.~Evans, M.~F.~Bode, T.~J.~O'Brien, \& M.~J.~Darnley}, 31

\bibitem[{{Bode}(2010)}]{Bode10}
{Bode}, M.~F. 2010, Astronomische Nachrichten, 331, 160

\bibitem[{{Bode} \& {Evans}(2008)}]{CNbook08}
{Bode}, M.~F. \& {Evans}, A. 2008, {Classical Novae, 2nd Edition}, ed. M.~F.
  {Bode} \& A.~{Evans}, Cambridge Astrophysics Series, No.~43, Cambridge:
  Cambridge University Press

\bibitem[{{della Valle} {et~al.}(1992){della Valle}, {Bianchini}, {Livio}, \&
  {Orio}}]{Della}
{della Valle}, M., {Bianchini}, A., {Livio}, M., \& {Orio}, M. 1992, \aap, 266,
  232

\bibitem[{{Downes} \& {Duerbeck}(2000)}]{Downes}
{Downes}, R.~A. \& {Duerbeck}, H.~W. 2000, \aj, 120, 2007

\bibitem[{{Drake} \& {Orlando}(2010)}]{Drake}
{Drake}, J.~J. \& {Orlando}, S. 2010, \apjl, 720, L195

\bibitem[{{Duerbeck}(1990)}]{Duerbeck}
{Duerbeck}, H.~W. 1990, in Lecture Notes in Physics, Berlin Springer Verlag,
  Vol. 369, IAU Colloq. 122: Physics of Classical Novae, ed. {A.~Cassatella \&
  R.~Viotti}, 34

\bibitem[{{Hanes}(1985)}]{Hanes}
{Hanes}, D.~A. 1985, \mnras, 213, 443

\bibitem[{{Harman} {et~al.}(2003){Harman}, {Bryce}, {O'Brien}, \&
  {Meaburn}}]{Dan}
{Harman}, D.~J., {Bryce}, M., {O'Brien}, T.~J., \& {Meaburn}, J. 2003, in IAU
  Symposium, Vol. 209, Planetary Nebulae: Their Evolution and Role in the
  Universe, ed. {S.~Kwok, M.~Dopita, \& R.~Sutherland}, 531

\bibitem[{{Helton} {et~al.}(2008){Helton}, {Woodward}, {Vanlandingham}, \&
  {Schwarz}}]{Helton}
{Helton}, L.~A., {Woodward}, C.~E., {Vanlandingham}, K., \& {Schwarz}, G.~J.
  2008, \iaucirc, 8989, 2

\bibitem[{{Hoard} {et~al.}(2002){Hoard}, {Wachter}, {Clark}, \&
  {Bowers}}]{Hoard}
{Hoard}, D.~W., {Wachter}, S., {Clark}, L.~L., \& {Bowers}, T.~P. 2002, \apj,
  565, 511

\bibitem[{{Hounsell} {et~al.}(2010){Hounsell}, {Bode}, {Hick}, {Buffington},
  {Jackson}, {Clover}, {Shafter}, {Darnley}, {Mawson}, {Steele}, {Evans},
  {Eyres}, \& {O'Brien}}]{Hounsell}
{Hounsell}, R., {Bode}, M.~F., {Hick}, P.~P., {et~al.} 2010, \apj, 724, 480

\bibitem[{{Iijima}(2002)}]{Iijima}
{Iijima}, T. 2002, \aap, 387, 1013

\bibitem[{{McLaughlin}(1945)}]{Mclaughlin}
{McLaughlin}, D.~B. 1945, \pasp, 57, 69

\bibitem[{{Munari} {et~al.}(2011{\natexlab{a}}){Munari}, {Ribeiro}, {Bode}, \&
  {Saguner}}]{Munari2}
{Munari}, U., {Ribeiro}, V.~A.~R.~M., {Bode}, M.~F., \& {Saguner}, T.
  2011{\natexlab{a}}, \mnras, 410, 525

\bibitem[{{Munari} {et~al.}(2011{\natexlab{b}}){Munari}, {Siviero},
  {Dallaporta}, {Cherini}, {Valisa}, \& {Tomasella}}]{Munari}
{Munari}, U., {Siviero}, A., {Dallaporta}, S., {et~al.} 2011{\natexlab{b}},
  \na, 16, 209

\bibitem[{{Payne-Gaposchkin}(1957)}]{PG}
{Payne-Gaposchkin}, C. 1957, {The galactic novae.}, ed. {Payne-Gaposchkin, C.},
  Amsterdam, North-Holland Pub.~Co.; New York, Interscience Publishers

\bibitem[{{Porter} {et~al.}(1998){Porter}, {O'Brien}, \& {Bode}}]{Porter}
{Porter}, J.~M., {O'Brien}, T.~J., \& {Bode}, M.~F. 1998, \mnras, 296, 943

\bibitem[{{Prialnik} \& {Kovetz}(1995)}]{Prialnik}
{Prialnik}, D. \& {Kovetz}, A. 1995, \apj, 445, 789

\bibitem[{{Rowles} \& {Froebrich}(2009)}]{Rowles}
{Rowles}, J. \& {Froebrich}, D. 2009, \mnras, 395, 1640

\bibitem[{{Schaefer}(2010)}]{Schaefer}
{Schaefer}, B.~E. 2010, \apjs, 187, 275

\bibitem[{{Shara}(1981)}]{Shara}
{Shara}, M.~M. 1981, \apj, 243, 268

\bibitem[{{Skrutskie} {et~al.}(1995){Skrutskie}, {Beichman}, {Capps},
  {Carpenter}, {Chester}, {Cutri}, {Elias}, {Elston}, {Huchra}, {Liebert},
  {Lonsdale}, {Monet}, {Price}, {Schneider}, {Seitzer}, {Stiening}, {Strom}, \&
  {Weinberg}}]{Skrutskie}
{Skrutskie}, M.~F., {Beichman}, C., {Capps}, R., {et~al.} 1995, in Bulletin of
  the American Astronomical Society, Vol.~27, Bulletin of the American
  Astronomical Society, 1392

\bibitem[{{Slavin} {et~al.}(1995){Slavin}, {O'Brien}, \& {Dunlop}}]{Slavin}
{Slavin}, A.~J., {O'Brien}, T.~J., \& {Dunlop}, J.~S. 1995, \mnras, 276, 353

\bibitem[{{Starrfield} {et~al.}(2008){Starrfield}, {Iliadis}, \&
  {Hix}}]{Starrfieldpaper}
{Starrfield}, S., {Iliadis}, C., \& {Hix}, W.~R. 2008, {in Classical Novae, 2nd
  Edition}, ed. {{Bode}, M.~F. and {Evans}, A.}, Cambridge Astrophysics Series,
  No.~43, Cambridge: Cambridge University Press, 77

\bibitem[{{Warner}(2008)}]{Warner}
{Warner}, B. 2008, {in Classical Novae, 2nd Edition}, ed. M.~F. {Bode} \&
  A.~{Evans}, Cambridge Astrophysics Series, No.~43, Cambridge: Cambridge
  University Press, 16

\bibitem[{{Weight} {et~al.}(1994){Weight}, {Evans}, {Naylor}, {Wood}, \&
  {Bode}}]{Weight}
{Weight}, A., {Evans}, A., {Naylor}, T., {Wood}, J.~H., \& {Bode}, M.~F. 1994,
  \mnras, 266, 761

\bibitem[{{Williams}(1992)}]{Williams}
{Williams}, R.~E. 1992, \aj, 104, 725

\bibitem[{{Yamaoka} {et~al.}(2008){Yamaoka}, {Itagaki}, {Nakano}, {Nevski},
  {Hsiao}, {Graham}, {Pritchet}, {Balam}, \& {Kazarovets}}]{Yamaokab}
{Yamaoka}, H., {Itagaki}, K., {Nakano}, S., {et~al.} 2008, \iaucirc, 8989

\end{thebibliography}

\end{document}